\documentstyle[aps]{revtex}
\begin{document}
\draft
\title{Temperature Variation of Ultra Slow Light in a Cold Gas}
\author{G. Morigi$^1$ and G.S. Agarwal$^{1,2}$}
\address{$^1$ Max Planck Institut f\"{u}r Quantenoptik, Garching 85748,
Germany.\\
$^2$ Physical Research Laboratory, Navrangpura, Ahmedabad 380009. India.}
\date{\today}
\maketitle
\begin{abstract}
A model is developed to explain the temperature dependence of the group velocity
as observed in the experiments of Hau et al (Nature {\bf397}, 594 (1999)). The
group velocity is quite sensitive to the change in the spatial density. The
inhomogeneity in the density and its temperature dependence are primarily
responsible for the observed behavior.
\end{abstract}


\newpage
\section{Introduction}

The phenomenon of Bose--Einstein condensation in atomic gases \cite{BEC,Review}
lends itself to the study of many fundamental effects. Among them, one aspect
presently being investigated both theoretically and experimentally is the interaction of light with atoms in the quantum
degeneracy regime \cite{Review}. In this context, the propagation of light inside a cold gas
is still an open problem. Because of the optical density, it is well known that the transmission 
of resonant light through a condensate is almost zero \cite{ZeroT}. However,
electromagnetically induced transparency (EIT) \cite{PhysToday} was found to allow the propagation of light by means of quantum coherence 
between different internal atomic levels \cite{GroupV,Kash}. In this context,   
Hau et al discovered a remarkable property of pulse propagation in a Bose condensate. These
authors demonstrated the slowing down of the group velocity of the pulse to 17 meters/sec
\cite{Nature}. Furthermore, they have shown a definite dependence of the group velocity on the 
temperature of the ultra-cold sample. One would like to understand the observed temperature
dependence from first principles. For this purpose, it is necessary
to extend the standard theory of EIT to a cold gas at finite temperature. However,
a theoretical description of this problem is rather complex. Complexities arise
when  one attempts a systematic treatment of interactions, finite temperature 
effects and dynamics. 
Most studies treat these aspects as disjoint: interactions are included in the 
zero temperature case to study the kinematical aspects \cite{Review,Elena}, whereas 
some dynamical aspects are studied using only the excitations within the electronic 
ground state \cite{Griffin}, and finite temperature effects are usually studied for noninteracting bosons 
\cite{Review,Scully,Thermo}.
A complete theory should study all these aspects together. However, a complete theory of 
the interaction of light and interacting particles is still unavailable, and a full 
numerical treatment is a rather hard task. Here, we present approximate but plausible 
arguments to explain the experimental observations in \cite{Nature}. 
The simplicity of our model allows for an analytical expression for the 
group velocity in the following cases: atoms confined in a box and by a 
harmonic  potential. We obtain results which reproduce the ones in 
\cite{Nature} for $T>T_c$. In particular, 
the treatment brings out the factors playing key roles in the 
phenomenon.
Here, we show that the variation of spatial density of atoms with temperature
is the major factor responsible for the temperature dependence of the  group
velocity. \\
The paper is organized as follows: In Section II the model is introduced. 
In Section III we derive the group velocity of a pulse propagating in an 
ideal gas confined inside a box, extend the calculation to the case of an ideal gas 
in a harmonic oscillator potential, and
present and discuss the results in relation to the experiment of Hau et al. 
In Section IV we present estimates for the group velocity in the interacting case
and in the limit of zero temperature.
   
\section{The Model}

In this Section we introduce the model used throughout this article. Here, 
we write the Maxwell-Bloch equations which describe the dynamics of the system 
consisting of light field and atoms. We derive the linear response of the medium
to a weak probe field, taking into account the quantum statistics of the atoms. 
The group velocity is then defined in the standard manner \cite{GroupRef}.

\subsubsection{Maxwell-Bloch Equations}

We consider a gas of $N$ non-interacting bosons. The relevant internal structure
corresponds to a three-level atom, with internal levels $|g\rangle$ (stable
state), $|r\rangle$, (metastable state) and $|e\rangle$ excited state, whose energies are
$\omega_g$, $\omega_r$ and $\omega_e$, respectively
(see Fig. 1). The radiative decay rate of the excited state is $\gamma=\gamma_g
+\gamma_r$, with $\gamma_g (\gamma_r)$ the rate of decay on the transition $|e\rangle
\to|g\rangle (|e\rangle\to|r\rangle)$.
Laser light with frequency $\omega_{lg}$ and wave vector $k_g$ drives the transition $|g\rangle\to|e\rangle$, 
whereas the transition $|r\rangle\to|e\rangle$ is driven by a field of frequency
$\omega_{lr}$ and wave vector $k_r$. 
The dynamics of the whole system is given by the Maxwell equation for the 
electric field vector ${\bf E}$

\begin{equation}
\nabla^2{\bf E}-\frac{1}{c^2}\frac{\partial^2}{\partial t^2}{\bf E}
=\frac{4\pi}{c^2}\frac{\partial^2}{\partial t^2}{\bf P}~,
\label{Maxwell}
\end{equation}
and by the optical Bloch equations for the density matrix equations of the $N$-atom gas.
For non-interacting atoms it suffices to consider the equations for the  one-atom
density matrix $\rho$, projected on the basis $\{ |j,\epsilon\rangle \}$ with 
$j=r,g,e$ and $|\epsilon\rangle$ the eigenvector
of the mechanical motion of one atom at the energy $\epsilon$. They have the form:

\begin{eqnarray}
& &\frac{\text{d}}{\text{d}t}\rho_{gg}(\epsilon,\epsilon')
=-i(\epsilon-\epsilon')\rho_{gg}(\epsilon,\epsilon') 
\label{Opt1}\\
& &+i\frac{\mbox{g}}{2}\sum_{\epsilon_1}
\left[  C_{\epsilon,\epsilon_1}^g\tilde{\rho}_{eg}(\epsilon_1,\epsilon') 
-\tilde{\rho}_{ge}(\epsilon,\epsilon_1) (C_{\epsilon',\epsilon_1}^g)^*\right]
+\gamma_g\rho_{ee}^g(\epsilon,\epsilon') ,
\nonumber\\
& &\frac{\text{d}}{\text{d}t}\rho_{rr}(\epsilon,\epsilon')
=-i(\epsilon-\epsilon')\rho_{rr}(\epsilon,\epsilon')
\label{Opt2}\\
& &+i\frac{\Omega}{2}\sum_{\epsilon_1}
\left[  C_{\epsilon,\epsilon_1}^r\tilde{\rho}_{er}(\epsilon_1,\epsilon')
-\tilde{\rho}_{re}(\epsilon,\epsilon_1) (C_{\epsilon',\epsilon_1}^r)^*\right]
+\gamma_r\rho_{ee}^r(\epsilon,\epsilon'),\nonumber\\
& &\frac{\text{d}}{\text{d}t}\rho_{ee}(\epsilon,\epsilon')
=-i(\epsilon-\epsilon'-i\gamma)\rho_{ee}(\epsilon,\epsilon')
\label{Opt3}\\
& &+i\frac{\mbox{g}}{2}\sum_{\epsilon_1}
\left[(C_{\epsilon_1,\epsilon}^g)^*\tilde{\rho}_{ge}(\epsilon_1,\epsilon')
-\tilde{\rho_{eg}}(\epsilon,\epsilon_1) C_{\epsilon_1,\epsilon'}^g\right] 
\nonumber\\
& &+i\frac{\Omega}{2}\sum_{\epsilon_1}\left[ 
(C_{\epsilon_1,\epsilon}^r)^*\tilde{\rho}_{re}(\epsilon_1,\epsilon')
-\tilde{\rho}_{er}(\epsilon,\epsilon_1) C_{\epsilon_1,\epsilon'}^r\right] ,
\nonumber\\
& &\frac{\text{d}}{\text{d}t}\tilde{\rho}_{ge}(\epsilon,\epsilon')
=-i(\epsilon-\epsilon'-\Delta_g^0-i\Gamma_{ge})
\tilde{\rho}_{ge}(\epsilon,\epsilon')\label{Opt4}\\
& &+i\frac{\mbox{g}}{2}\sum_{\epsilon_1}
\left[ C_{\epsilon,\epsilon_1}^g\rho_{ee}(\epsilon_1,\epsilon')
-\rho_{gg}(\epsilon,\epsilon_1)C_{\epsilon_1,\epsilon'}^g\right] 
\nonumber\\
& &-i\frac{\Omega}{2}\sum_{\epsilon_1}
\tilde{\rho}_{gr}(\epsilon,\epsilon_1)C_{\epsilon_1,\epsilon'}^r ,\nonumber\\
& &\frac{\text{d}}{\text{d}t}\tilde{\rho}_{re}(\epsilon,\epsilon')
=-i(\epsilon-\epsilon'-\Delta_r^0-i\Gamma_{re})\tilde{\rho}_{re}(\epsilon,\epsilon')
\label{Opt5}\\
& &+i\frac{\Omega}{2}\sum_{\epsilon_1}
\left[ C_{\epsilon,\epsilon_1}^r\rho_{ee}(\epsilon_1,\epsilon')
-\rho_{rr}(\epsilon,\epsilon_1)C_{\epsilon_1,\epsilon'}^r\right]
\nonumber\\
& &-i\frac{\mbox{g}}{2}
\sum_{\epsilon_1}\tilde{\rho}_{rg}(\epsilon,\epsilon_1)C_{\epsilon_1,\epsilon'}^g ,
\nonumber\\
& &\frac{\text{d}}{\text{d}t}\tilde{\rho}_{gr}(\epsilon,\epsilon')
=-i(\epsilon-\epsilon'-\left(\Delta_g^0-\Delta_r^0\right)
-i\Gamma_{gr})\tilde{\rho}_{gr}(\epsilon,\epsilon')\label{Opt6}\\
& &+i\frac{\mbox{g}}{2}\sum_{\epsilon_1}
C_{\epsilon,\epsilon_1}^g\tilde{\rho}_{er}(\epsilon_1,\epsilon')
-i\frac{\Omega}{2}\sum_{\epsilon_1}
\tilde{\rho}_{ge}(\epsilon,\epsilon_1)(C_{\epsilon',\epsilon_1}^r)^* ,
\nonumber
\end{eqnarray}
where $\rho_{ij}(\epsilon,\epsilon')=\langle i,\epsilon|\rho|j,\epsilon'\rangle$ 
($i,j=r,e,g$), 
$C_{\epsilon,\epsilon'}^j=\langle \epsilon|\exp(i{\bf k_j}\cdot{\bf r})|\epsilon'\rangle$, 
$\tilde{\rho}_{ej}=\rho_{ej}\mbox{e}^{-i\omega_{lj}t}$,
$\tilde{\rho}_{rg}=\rho_{rg}\mbox{e}^{-i(\omega_{lg}-\omega_{lr})t}$ and 
$\tilde{\rho}_{ij}=(\tilde{\rho}_{ji})^*$ for $i\neq j$.
Here, $c$ is the speed of light, ${\bf P}$ is the polarization of the medium, 
$\Delta_j^0=\omega_{e}-\omega_{j}-\omega_{lj}$ ($j=g,r$) is the detuning. Rabi
couplings are given by 
$g=|{\bf d_{eg}}\cdot {\bf E}|/\hbar$ and $\Omega=|{\bf d_{er}}\cdot{\bf E}|/\hbar$, 
where ${\bf d_{ej}}$ is the dipole moment of the transition 
$|e\rangle\to|j\rangle$. Finally, $\rho_{ee}^j(\epsilon,\epsilon')$ describes the density 
matrix after a spontaneous emission event on the transition $|e\rangle\to |j\rangle$:

\begin{equation}
\label{feed}
\rho_{ee}^j(\epsilon,\epsilon')=\frac{3}{8\pi}
\sum_{l=1,2}\sum_{\epsilon_1,\epsilon_2}\int\text{d}\Omega_{\bf \hat{k}}
|{\bf \hat{d}_{je}}\cdot\alpha^l({\bf \hat{k}})|^2
\langle \epsilon|\text{e}^{i{\bf k_{j}}\cdot {\bf r}}
|\epsilon_1\rangle\rho_{ee}(\epsilon_1,\epsilon_2)
\langle \epsilon_2|\text{e}^{-i{\bf k_{j}}\cdot {\bf r}}|\epsilon'\rangle~~,
\end{equation}
where $\alpha^1({\bf \hat{k}})$ and $\alpha^2({\bf \hat{k}})$ form a set of polarization vectors
orthogonal to ${\bf \hat{k}}$, and ${\bf \hat{d}_{je}}={\bf d_{je}}/|{\bf d_{je}}|$.\\
In deriving the above equations, we made the rotating wave approximation and
transformed to a reference frame rotating at the optical frequency of the laser.
Furthermore, in Eqs. (\ref{Opt1})-(\ref{Opt6}) we have introduced the loss-rates 
$\Gamma_{ij}$, which take into account the effects of other mechanisms of decoherence.
In the ideal case $\Gamma_{ge}=\Gamma_{re}=\gamma/2$, whereas $\Gamma_{gr}=0$.

\subsubsection{Susceptibility and Group velocity}

In an isotropic medium, the linear susceptibility $\chi$ is defined by the expression 
\cite{Delone}

\begin{equation}
{\bf P}(t)=\int_{-\infty}^t \text{d}t'\chi(t-t'){\bf E}(t')~.
\label{susc1}
\end{equation}
Assuming that two light fields are propagating through the medium along the
$\hat{z}$-direction, we can write the electric field $E({\bf r},t)$ and the 
atomic polarization $P(\bf{r},t)$ as $E({\bf r},t)=\sum_{j=g,r}
[E_0^j({\bf r},t)\exp(+ik_jz-i\omega_{lj} t)+c.c.]/2$ and 
$P({\bf r},t)=\sum_{j=g,r}
[P_0^j({\bf r},t)\exp(+ik_jz-i\omega_{lj} t)+c.c.]/2$ where $E_0^j$, $P_0^j$ 
are respectively the slowly-varying envelopes of the electric field and 
atomic polarization at frequency $\omega_{lj}$. The Fourier transform of Eq. (\ref{susc1})
gives $P_0(\omega)= E_0(\omega)\chi(\omega)$. 
From the relation ${\bf P}(t)=\mbox{Tr}\{\rho^N(t) {\bf d}\}$ for the polarization
with ${\bf d}$ the
atomic dipole moment operator, we find that the macroscopic polarization of the medium 
at the position ${\bf r}$ is
\begin{equation}
P_0^g({\bf r},\omega_{lg})/2=\langle {\bf r}|\rho^N_{eg}d_{ge}\mbox{e}^{-ik_gz}|{\bf r} \rangle~,
\label{EqCohe}
\end{equation}
where $\rho^N$ is the $N$-atom density matrix, that has the following form in
the energy representation.

\begin{equation}
\rho_{eg}^N=\sum_{\epsilon,\epsilon'}\rho_{eg}^N(\epsilon,\epsilon')
|\epsilon\rangle\langle\epsilon'|~.
\end{equation}
The N-atom optical coherence density matrix has to be obtained from the solution of 
Eqs. (\ref{Opt1})-(\ref{Opt6}) subject to the initial condition:

\begin{equation}
\label{Equilibrium}
\rho^N(0)=\sum_{\epsilon}N(\epsilon)|g,\epsilon\rangle\langle g,\epsilon| ,
\label{Initio}
\end{equation}
where $N(\epsilon)$ is the number of atoms in the ground state with energy $\epsilon$.
The Eqs. (\ref{Opt1})-(\ref{Opt6}) are to be solved to first order in the field $E_0^g$ and 
to all orders in the field $E_0^r$. In this work we are interested in the steady state of the atoms
with the field, which is assumed to be reached on a time-scale much shorter than the 
thermalization time-scale of the gas. On the basis of this hypothesis, we assume that the
initial condition (\ref{Initio}) and Eqs. (\ref{Opt1})-(\ref{Opt6})
determine the steady state solution. Note that the coefficients $C_{\epsilon,\epsilon'}$ 
determine the one-atom energy states involved in the transition induced by the laser field.
For free bosons these coefficients have a simple form, as it can be seen in
section III.A. For harmonic oscillator potentials the coefficients 
$C_{\epsilon,\epsilon'}$ are given in terms
of Laguerre polynomials, where the number of vibrational states which are coupled depends on the 
ratio between the recoil frequency over the trap frequency.
In section III.B we use an approximate treatment for this case.\\
Once the susceptibility is known, the dispersion relation of light
in the medium is given \cite{Delone} and we can evaluate the group velocity, defined as $v_g
=\frac{\partial \omega}{\partial k_g}|_{\omega=\omega_{eg}}$. In the limit $N\chi\ll 1$
the group velocity has the form:

\begin{equation}
v_g =\frac{c}
{1+2\pi\chi'|_{\omega=\omega_{lg}}
+2\pi\omega_{lg}\frac{\partial \chi'}{\partial \omega}|_{\omega=\omega_{lg}}}~,
\label{GroupV}
\end{equation}
where $\chi'=\text{Re}(\chi)$.

\section{Evaluation of the group velocity}

In this section we derive an analytical expression for the group velocity of
a laser pulse propagating through an ideal gas of ultracold atoms. We investigate 
two cases: atoms in a box and atoms confined by a harmonic 
oscillator potential. Finally, we do the numerical calculations for the case of a gas of
sodium atoms and discuss the results in relation with the experimental
data of \cite{Nature}.

\subsection{Group velocity in a gas of free non-interacting bosons}

We consider a gas of $N$ bosons in a box of volume $V$. In this case the atomic
wave vector eigenstates $|{\bf k}\rangle$ are also energy eigenstates with
eigenvalues $\epsilon =\frac{\hbar^2{\bf k}^2}{2m}$, where $m$ is the atomic 
mass. So, we project Eqs. (\ref{Opt1})-(\ref{Opt6}) on the motional basis $\{|{\bf k}\rangle\}$. Then
the coefficients $ C^j_{\epsilon\epsilon'}$ appearing in the density matrix
equations have the form $ C^j_{\epsilon\epsilon'}\equiv
C^j_{{\bf k},{\bf k'}}=\delta_{{\bf k},{\bf k}+{\bf k_j}}$. 
We substitute these values into Eqs. (\ref{Opt1})-(\ref{Opt6})
and solve the equations in the steady state limit.
To first order in $\mbox{g}/\Omega$ and $\mbox{g}/\Gamma$, and assuming that
at $t=0$ the gas is in thermal equilibrium, 
the steady-state optical coherence $\rho_{eg}$ is found to be 
\begin{eqnarray}
\rho_{eg}({\bf k}-{\bf k_g},{\bf k})
&=&\mbox{g}\frac{2i\left(\Gamma_{gr}+i(\Delta_g-\Delta_r)\right)}
{\Omega^2+4\left(\Gamma_{ge}+i\Delta_g\right)\left(\Gamma_{gr}+i(\Delta_g-\Delta_r)\right)}
\nonumber\\
&=&\frac{\mbox{g}}{2\Gamma_{ge}}\frac{1}{\frac{\Delta_g}{\Gamma_{ge}}-i
-i\frac{\Omega^2/4}{\Gamma_{ge}\left(\Gamma_{gr}+i(\Delta_g-\Delta_r)\right)}}~~,
\label{OptCoh}
\end{eqnarray}
where $\Delta_j$ is the detuning defined as 
\begin{equation}
\Delta_j=\tilde{\Delta}_j^{0}+\frac{\hbar {\bf k}_j\cdot {\bf k}}{m}~~,
\label{Det}
\end{equation}
with $\tilde{\Delta}_j^{0}=\Delta_j^{0}+\omega_R$, and $\omega_R$  is the recoil 
frequency defined as $\omega_R=\hbar k^2/2m$. 
Using (\ref{OptCoh}) in (\ref{EqCohe}), we find the expression for the susceptibility 

\begin{equation}
\chi_{ge}(\omega_{lg})
=\chi^0\sum_{\bf k} N({\bf k})\frac{1}{V}
\frac{1}{\frac{\Delta_g}{\Gamma_{ge}}-i
-i\frac{\Omega^2/4}{\Gamma_{ge}\left(\Gamma_{gr}+i(\Delta_g-\Delta_r)\right)}}~~.
\label{susce}
\end{equation}
The sum in Eq. (\ref{susce})
is over all the motional states weighted by their statistical 
occupation $N({\bf k})$,

\begin{equation}
N({\bf k})=\frac{1}{f^{-1}\exp\left(\beta\hbar^2k^2/2m\right)-1}~~, 
\end{equation}
where $f$ is the fugacity, $\beta=1/K_B T$ and $T$ is the temperature, and $k=|{\bf k}|$.
Here, $\chi^0$ is the one-atom susceptibility, defined as

\begin{equation}
\chi^0=\frac{\bf
|d_{ge}|^2}{\Gamma_{ge}\hbar}\equiv\frac{3\lambda^3}{32\pi^3}~~,
\end{equation}
where $\lambda$ is the optical wavelength of the transition $g\to e$, 
$\lambda=2\pi c/\omega_{ge}$, and $\Gamma_{ge}=\gamma/2$.
In the following, we assume that $|{\bf k}_g-{\bf k}_r|\ll k_g,k_r,\Gamma_{gr}$. Therefore,
the dependence on $k$ in the denominator of Eq. (\ref{susce})
is mainly due to the first term $\Delta_g$, and we may rewrite Eq. (\ref{susce}) as

\begin{equation}
\chi_{ge}(\omega_{lg})=\frac{\chi^0}{V}\sum_{\bf k} N({\bf k})
\frac{1}{\frac{\hbar k_g}{m\Gamma_{ge}}k_z-\zeta}~~ ,
\label{susce11}
\end{equation}
where $k_z={\bf k}\cdot\hat{z}$ and $\zeta$ is a complex number independent
of $k$;

\begin{equation}
\zeta=-\frac{\tilde{\Delta}_g^{0}}{\Gamma_{ge}}
+i+i\frac{\Omega^2/4}{\Gamma_{ge}\left(\Gamma_{gr}+i(\Delta_g^{0}-\Delta_r^{0})\right)}~~.
\label{Zeta}
\end{equation}
After evaluating the susceptibility as given by Eq. (\ref{susce11}), the group velocity 
can be found using Eq. (\ref{GroupV}). \\
In the following, we investigate the behaviour of the cloud close to the critical
point, dividing our investigation into two regimes: above and below the critical temperature
$T_c$.

\subsubsection{Above the critical temperature}

Above the critical temperature and in the limit of large volumes, 
one can replace the sum in Eq. (\ref{susce11}) with an integral in three-dimensions
\cite{Huang}. The expression to evaluate is now

\begin{eqnarray}
\label{susce0free}
& &{\chi_{ge}(\omega_{lg})}_+
=\frac{\chi^0}{(2\pi)^3}
\int_{-\infty}^{\infty}\text{d}k_x\int_{-\infty}^{\infty}\text{d}k_y
\int_{-\infty}^{\infty}\text{d}k_z \label{inte1}\\ 
& &\cdot\left[\frac{1}{f^{-1}\exp\left(\beta\hbar^2k^2/2m\right)-1}\right]
\left[\frac{1}{\frac{\hbar k_g}{m\Gamma_{ge}}k_z-\zeta}\right]~~,
\nonumber
\end{eqnarray}
where for convenience we have chosen to integrate in the cartesian coordinates.
We write $N({\bf k})$ as 

\begin{equation}
N({\bf k})=\sum_{l=1}^{\infty}f^l\exp\left(-l\beta\hbar^2k^2/2m\right)~~,
\end{equation}
and use it in Eq. (\ref{inte1}) to obtain

\begin{eqnarray}
& &{\chi_{ge}(\omega_{lg})}_+
=  \frac{\chi^0}{(2\pi)^3}\sum_{l=1}^{\infty}z^l
\int_{-\infty}^{\infty}\text{d}k_z 
\frac{\exp\left(-l\beta\hbar^2k_z^2/2m\right)}{\frac{\hbar k_g}{m\Gamma_{ge}}k_z-\zeta}
\\
& &\cdot
\int_{-\infty}^{\infty}\text{d}k_x\exp\left(-l\beta\hbar^2k_x^2/2m\right)
\int_{-\infty}^{\infty}\text{d}k_y\exp\left(-l\beta\hbar^2k_y^2/2m\right)~~. 
\nonumber
\end{eqnarray}
This can be written in terms of the standard functions

\begin{equation}
{\chi_{ge}(\omega_{lg})}_+
=i\frac{\chi^0}{8A\pi}
\left(\frac{2mK_BT}{\hbar^2}\right)^{3/2} \sum_{l=1}^{\infty}\frac{f^l}{l}
w\left(\sqrt{l}\frac{\zeta}{A}\right)~~,
\label{equ1}
\end{equation}
where

\begin{equation}
A=\sqrt{\frac{2K_BT}{m}}\frac{k_g}{\Gamma_{ge}}~~,
\label{Adef}
\end{equation}
and where the function $w$ is defined as

\begin{equation}
w(x)=\exp(-x^2)\left(\text{erf}(ix)+1\right)~~.
\end{equation}
Given the critical temperature for an ideal Bose gas 

\begin{equation}
T_c=\frac{2\pi\hbar^2}{mK_B}\left(\frac{n}{g_{3/2}(1)}\right)^{2/3}~~,
\end{equation}
where $n(=N/V)$ is the density of atoms, we can rewrite Eq. (\ref{equ1}) as

\begin{equation}
\chi_{ge}(\omega_{lg})_{+}=in\chi^{0}\frac{T}{T_c}\frac{1}{g_{3/2}(1)A_c}
\sum_{l=1}^{\infty}\frac{f^l}{l} w\left(\sqrt{l}\frac{\zeta}{A}\right)~~,
\label{susce13}
\end{equation}
with

\begin{equation}
A_c=2\frac{k_{g}}{\Gamma_{ge}}\frac{\hbar}{m}\left(\frac{n}{g_{3/2}(1)}\right)^{1/3}
=\sqrt{\pi} A|_{T=T_c}~~,
\end{equation}
and 

\begin{equation}
\frac{A}{A_c\sqrt{\pi}}=\sqrt{\frac{T}{T_c}}~~.
\end{equation}

\subsubsection{Below the critical temperature}

For $T<T_c$, we use the expression for the ground state population in the 
thermodynamic limit \cite{Huang}, to obtain

\begin{equation}
\label{Susce0free1}
\chi_{ge}(\omega_{lg})_{-}={\chi_{ge}(\omega_{lg})_{+}}|_{f=1}
-\frac{\chi^0}{\zeta}n\left[1-\left(\frac{T}{T_c}\right)^{3/2}\right]~~,
\end{equation}
where the second term on the RHS describes the contribution of the condensed
phase.

\subsubsection{Regime of parameters and approximations}

For currently studied optical transitions the argument of the $w$ function in 
Eq. (\ref{susce13}) is $y=\sqrt{l}\zeta/A\gg 1$ for any value of $l\ge 1$. 
Therefore, the asymptotic expansion of the $w$ function can be applied \cite{Abramo}:

\begin{equation}
w(y)=\frac{i}{\sqrt{\pi}y}+\frac{i}{2\sqrt{\pi}y^3}~.
\label{ExpAbramo}
\end{equation}
We substitute this expansion into Eq. (\ref{susce13}), and obtain for $T>T_c$

\begin{eqnarray}
\chi_{ge}(\omega_{lg})_{+}
&=&-n\chi^{0}\frac{T}{T_c}\frac{A}{\sqrt{\pi}A_c}\frac{1}{g_{3/2}(1)\zeta}
\sum_{l=1}^{\infty}\frac{f^l}{l}\left[\frac{1}{\sqrt{l}}+\frac{A^2}{\sqrt{l^3}\zeta^2}
\right]
\nonumber\\
&=&-n\chi^{0}\left(\frac{T}{T_c}\right)^{3/2}\frac{1}{g_{3/2}(1)\zeta}\left[g_{3/2}(f)
+g_{5/2}(f)\frac{A^2}{\zeta^2}\right]\nonumber\\
&=&-n\frac{\chi^{0}}{\zeta}\left[1+\left(\frac{T}{T_c}\right)^{3/2}
\frac{g_{5/2}(f)}{g_{3/2}(1)}\frac{A^2}{\zeta^2}\right]~~,
\label{susce14}
\end{eqnarray} 
where we have used the relation $g_{3/2}(f)/g_{3/2}(1)=T_c/T$ for $T>T_c$. For
$T<T_c$ the susceptibility will now have the form

\begin{equation}
\chi_{ge}(\omega_{lg})_{-}
= -n\frac{\chi^{0}}{\zeta}\left[1+\left(\frac{T}{T_c}\right)^{3/2}
\frac{g_{5/2}(1)}{g_{3/2}(1)}\frac{A^2}{\zeta^2}\right]~.
\label{susce15}
\end{equation}
Using Eqs. (\ref{susce14}) and (\ref{susce15})in the formula (13) we
find the group velocity. Note that the dependence on the temperature comes
in at higher order in the expansion $A/\zeta$. Clearly a significant temperature
dependence for a free gas can come only for narrow optical transitions.

\subsection{Group velocity in a gas of trapped non-interacting bosons}

Let us now consider a cloud of atoms trapped in a three-dimensional harmonic 
potential with cylindrical symmetry, so that the one-atom Hamiltonian 
describing the mechanical motion has the form

\begin{equation}
H=\frac{{\bf p}^2}{2m}+V({\bf r})~~,
\end{equation}
where $V({\bf r})$ is the harmonic oscillator potential in cylindrical 
coordinates

\begin{equation}
V({\bf r})=\frac{1}{2}m\left(\nu_r^2 r^2+\nu_z^2z^2\right)~~,
\end{equation}
with $\nu_r$, $\nu_z$ trap frequencies in the radial and axial directions, respectively.
In order to evaluate the susceptibility in the steady state, we solve 
 Eqs. (\ref{Opt1})-(\ref{Opt6})
in the semiclassical limit for the atomic motion, and we sum over the states using the
semiclassical statistical distribution \cite{Review,Jesus}. This limit is valid
when treating the non-condensed fraction of atoms for temperatures $T$ fulfilling 
the condition $K_BT\gg \hbar\nu$, and under the condition $\Gamma,\Delta\gg\nu$.
The hypothesis is justified in the range of parameters of \cite{Nature} and
simplifies considerably the treatment, allowing for an analytical solution
of the group velocity. Then the coeffients $C_{\epsilon,\epsilon'}$ simplify
to their semiclassical values $\langle C_{\epsilon,\epsilon'}\rangle\approx
\delta_{{\bf p},{\bf p'}+\hbar{\bf k}}
\delta_{{\bf r},{\bf r'}}$, where ${\bf p}$, ${\bf r}$ are now the classical canonical
coordinates of a harmonic oscillator with energy $E={\bf p}^2/2m+V({\bf r})$.
In this limit, the optical coherence $\rho_{eg}$ appearing in Eq. (\ref{EqCohe})
has the form:

\begin{equation}
\rho_{eg}({\bf r},{\bf p})
=\frac{\mbox{g}}{2\Gamma_{ge}}\frac{1}{\frac{\Delta_g}{\Gamma_{ge}}-i
-i\frac{\Omega^2/4}{\Gamma_{ge}\left(\Gamma_{gr}+i(\Delta_g-\Delta_r)\right)}}~~,
\end{equation}
with $\Delta_j$ defined in Eq. (\ref{Det}), and the susceptibility is given by the 
expression:

\begin{eqnarray}
\chi_{ge}({\bf r},\omega_{lg})
&=&{\chi^0}\int\frac{\text{d}^3{\bf p}}{(2\pi\hbar)^3}N({\bf r},{\bf p})\nonumber\\
& &\frac{1}{\frac{\tilde{\Delta}_g}{\Gamma_{ge}}-i
-i\frac{\Omega^2/4}{\Gamma_{ge}\left(\Gamma_{gr}+i(\Delta_g-\Delta_r)\right)}}~~,
\label{susceHO}
\end{eqnarray}
where the semiclassical statistical distribution is 

\begin{equation}
N({\bf r},{\bf p})=\frac{1}{f^{-1}\exp(\beta(\frac{p^2}{2m}+V({\bf r}))-1}~~.
\end{equation}
When considering the condensate contribution to the optical susceptibility,
one should evaluate $\rho_{eg}(\epsilon ,\epsilon')$ and sum over the final
states with energy $\epsilon$. However, in the regime
$\omega_R/\nu\gg 1$ we may apply the semiclassical approximation
to the final states. The final semiclassical energy is the recoil energy, and we 
can write the optical coherence for the condensate contribution as
    
\begin{equation}
\rho_{eg}
=\frac{\mbox{g}}{2\Gamma_{ge}}
\frac{1}{\frac{\tilde{\Delta}_g^0}{\Gamma_{ge}}-i
-i\frac{\Omega^2/4}{\Gamma_{ge}\left(\Gamma_{gr}+i(\Delta_g^0-\Delta_r^0)\right)}}~~.
\label{rhoCon}
\end{equation}
The ground state occupation is given in the thermodynamic limit by 
\begin{equation}
N^{(0)}=N\left(1-\left(\frac{T}{T_c}\right)^3\right) 
\end{equation}
where $T_c$ is the  critical temperature of the trapped gas, $K_BT_c=
\hbar(\nu_z\nu_r^2)^{1/3}(N/g_3(1))^{1/3}$.\\
Contrary to the case of free bosons, 
the group velocity is not directly given by the formula (\ref{GroupV}), because of 
the spatial variation of the atomic density and therefore of the susceptibility. 
Here, we evaluate the group velocity
using a method equivalent to the experimental one of \cite{Nature}, 
i.e. we estimate the size $D$ of the cloud and 
calculate the delay $\Delta t$ of a pulse 
propagating across a selected region of a cold gas with respect to a pulse propagating in 
the vacuum. The group velocity is then given by the ratio of the size over the delay 
$v_g^{\mbox{\small{exp}}}=D/\Delta t$.
Assuming that the light is propagating along the
$\hat{z}$-axis and cuts a cylinder inside of the volume with section $S$ and centered on the 
$\hat{z}$-axis of the cloud, we write the delay $\langle\Delta t\rangle$ as
the average over the section $S$ of all the delays $\Delta t(r)$ of pulses propagating 
at distance $r$ from the axis of the cloud

\begin{equation}
\langle \Delta t\rangle=\frac{1}{\pi R^2}\int_0^R \text{d}r2\pi r \Delta t(r)~~,
\label{Ave1}
\end{equation} 
where $R$ is the radius of the illuminated circular section $S$ of the cloud and 
$\Delta t(r)$ is defined as

\begin{equation}
\Delta t(r)=\int_{-L(r)}^{L(r)}\text{d}z [v_g(r,z)]^{-1}~~,
\label{Ave2}
\end{equation}
where $v_g$ is defined in Eq. (\ref{GroupV}), and 
$L(r)$ is half the length of the path along the cloud. Note that $R$ will, in
principle, depend on the size of the incoming Gaussian beam. However, 
in the experiment of \cite{Nature}, $R$ is the radius
of a pinhole set before the measuring apparatus. The delay time is experimentally
obtained by measuring the difference between the delay time of the pulse propagating
across the cloud and the one of a pulse propagating in the vacuum. Assuming that 
$L(r)=L$  is the distance between a slit before the cloud and the photomultiplier, 
the final delay will be $\Delta t=\langle \Delta t\rangle-L/c$.
In the following, we evaluate the group velocity as a function of the 
temperature above and below the critical temperature.

\subsubsection{Above the critical temperature}

The integral over the momenta in Eq. (\ref{susceHO}) can be evaluated along the lines of
the procedure outlined in Eqs. (\ref{susce0free})-(\ref{equ1}). One finds

\begin{eqnarray}
& &\chi_{ge}({\bf r},\omega_{lg})_{+}
\\
&=&i\pi^2\chi^0\frac{(2m K_BT)^{3/2}}{(2\pi\hbar)^3}\frac{1}{A}
\sum_{l=1}^{\infty}\frac{f^l}{l}
w\left(\sqrt{l}\frac{\zeta}{A}\right)\text{e}^{-l\beta V(r)}~~.
\nonumber
\end{eqnarray}
Again, the considerations on the $w$ function made in the free case are applicable, 
and using its asymptotic expansion [Eq.(\ref{ExpAbramo})], one gets

\begin{eqnarray}
&&\chi_{ge}({\bf r},\omega_{lg})_{+}
=-\frac{\chi^0}{\zeta}\left(\frac{m K_BT}{2\pi\hbar^2}\right)^{3/2}\\
&&\cdot\left(
g_{3/2}\left(f\text{e}^{-\beta V({\bf r})}\right)
+g_{5/2}\left(f\text{e}^{-\beta V({\bf r})}\right)\frac{A^2}{\zeta^2}\right)~~.
\nonumber
\end{eqnarray}
For $L\gg D_z(T)$, where $D_z(T)$ is the axial thermal size of the cloud,\
we can replace $L$ by $\infty$ in the integral (\ref{Ave2}). Therefore,
the delay of a beam propagating along the z--axis, and 
entering the cloud at a distance $r$ from the cloud axis is 

\begin{eqnarray}
&&\Delta t(r)_{+}
=-2\pi\frac{\omega}{c}\frac{\partial\zeta}{\partial\Delta}
\frac{\chi^0}{\zeta^2}\frac{m(K_BT)^2}{2\pi\hbar^3\nu_z}
\sum_{l=1}^{\infty}\frac{f^l}{l^{2}}\\
&&\left(1+3\frac{A^2}{\zeta^2}\frac{1}{l}\right)
\text{e}^{-l\beta m\nu_r^2r^2/2}~.
\nonumber
\end{eqnarray}

Above the critical temperature the total delay $\langle\Delta t\rangle_{+}$
is thus

\begin{equation}
\label{DtHO}
\langle \Delta t\rangle_+
=  -2\pi\frac{\omega}{c}\frac{\partial\zeta}{\partial\Delta}
\frac{\chi^0}{\zeta^2}\frac{(K_BT)^3}{\hbar^3\nu_z\nu_r^2}
\frac{2}{\pi R^2}
\sum_{l=1}^{\infty}\frac{f^l}{l^{3}}
\left(1+3\frac{A^2}{\zeta^2}\frac{1}{l}\right)\left[1-\text{e}^{-l\beta
m\nu_r^2R^2/2}\right].
\end{equation}
Here, we take the size of the cloud to be the variance of the thermal distribution along
the $\hat{z}$-axis, and 
thus $D_z=\sqrt{2K_BT/m\nu_z^2}$. The group velocity above the critical temperature is
then $D_z/\langle \Delta t\rangle_+$. Note that, in the limit $R\ll D_r(T)$,
where
$D_r(T)=\sqrt{2K_BT/m\nu_r^2}$ is the radial thermal size of the cloud, the exponential appearing in
Eq. (\ref{DtHO}) can be expanded to yield $\langle \Delta t\rangle_+\approx 1/T$. 
Since $D_z(T)\propto \sqrt{T}$, the group velocity 
depends on the temperature as $v_g\propto T^{3/2}$. The same behaviour can be found 
when considering the other limiting case, i.e. $R\approx D_r(T)/\sqrt{2}$,
corresponding to averaging over the whole cloud. Then, the delay time has the form:

\begin{eqnarray}
\langle \Delta t\rangle_+
&=    &-2\pi\frac{\omega}{c}\frac{\partial\zeta}{\partial\Delta}
\frac{\chi^0}{\zeta^2}\frac{(K_BT)^3}{\hbar^3\nu_z\nu_r^2}
\frac{2}{\pi R^2}
\sum_{l=1}^{\infty}\frac{f^l}{l^{3}}
\left(1+3\frac{A^2}{\zeta^2}\frac{1}{l}\right)
\nonumber\\
&=    &-2\pi\frac{\omega}{c}\frac{\partial\zeta}{\partial\Delta}
\frac{\chi^0}{\zeta^2}N\left(\frac{T}{T_c}\right)^3\frac{2}{\pi R^2}
\left[\frac{g_3(f)}{g_3(1)}+3\frac{A^2}{\zeta^2}
\frac{g_4(f)}{g_3(1)}\right]
\nonumber\\
&=    &-2\pi\frac{\omega}{c}\frac{\partial\zeta}{\partial\Delta}
\frac{\chi^0}{\zeta^2}N\frac{2}{\pi R^2}
\left[1+3 \left(\frac{T}{T_c}\right)^3
\frac{A^2}{\zeta^2}\frac{g_4(f)}{g_3(1)}\right]
\label{HOtotal}
\end{eqnarray}
where we have used the definition of critical temperature and the relation
$g_3(1)/g_3(f)=(T/T_c)^3$ for $T>T_c$. From Eq. (\ref{HOtotal}) one sees that
the dependence of the delay time on the temperature 
of the sample appears principally in the spatial-average term $1/R^2$, which is proportional
to $1/T$. Thus, the main dependence of the group velocity on the temperature comes in 
through the volume of the cloud, since $v_g= D_z(T)/\langle \Delta t\rangle_+
\propto D_z(T) R^2\propto T^{3/2}$, and the variation
of the group velocity with temperature is mainly due to the change of volume of the cloud.

\subsubsection{Below the critical temperature}

For the ground state of the harmonic potential,
we use the expression for the ground state population in the 
thermodynamic limit \cite{Review} and the optical coherence as given in 
Eq. (\ref{rhoCon}) to obtain

\begin{eqnarray}
\chi_{ge}({\bf r},\omega_{lg})_{-}
&=&\chi_{ge}({\bf r},\omega_{lg})_{+}|_{f=1}
\\
&-&\chi^0\frac{1}{\zeta}N|\phi({\bf r})|^2
\left(1-\left(\frac{T}{T_c}\right)^3\right),
\nonumber
\end{eqnarray}
where $N$ is the total number of particles, and $\phi({\bf r})$ is the harmonic 
oscillator ground state wavefunction. The delay $\langle \Delta t\rangle_-$ is 
given by 

\begin{equation}
\langle \Delta t\rangle_-=\langle \Delta t\rangle_+|_{f=1}+
\langle \Delta t\rangle_C ,
\end{equation}
where $\langle\Delta t\rangle_C$ is the contribution to the total delay given
by the pulses which cross the condensate

\begin{equation}
\langle \Delta t\rangle_C=-2\pi\frac{\omega}{c}
\frac{\partial\zeta}{\partial\Delta}
\frac{\chi^0}{\zeta^2}N\left(1-\left(\frac{T}{T_c}\right)^3\right)F_C.
\end{equation}
Here $F_C$ is the average of the ground state wavefunction, which, according to Eqs. 
(\ref{susceHO}), (\ref{Ave1}), (\ref{Ave2}) is

\begin{equation}
F_C
=\frac{2}{R^2}\int_0^{R}r\text{d}r\frac{\text{e}^{-r^2/a_{0r}^2}}
{\pi^{3/2}a_{0r}^2 a_{0z}}\int_{-L}^{L}\text{d}z
\text{e}^{-z^2/a_{0z}^2}\approx\frac{2}{\pi R^2},
\end{equation}
where $a_{0j}=\sqrt{\hbar/m\nu_j}$ is the size of the ground state of the
harmonic oscillator in the $j$ direction ($j=r,z$).  
In order to evaluate the size of the cloud, we consider that for $T<T_c$, a 
fraction $(T/T_c)^3$ of the atoms is outside of the condensate, whereas a 
fraction $1-(T/T_c)^3$ is in the condensate. Applying the
semiclassical approximation to the non-condensate part, we have that
\begin{equation}
\langle z^2\rangle=(T/T_c)^3\langle z^2\rangle_{NC}+[1-(T/T_c)^3]\langle
z^2\rangle_C ,
\end{equation}
which leads us to defining the size of the cloud to be:
\begin{equation}
D_z=\sqrt{2}\left[\left(\frac{T}{T_c}\right)^3 R_+^2 +
\left(1-\left(\frac{T}{T_c}\right)^3\right)a_{0z}^2\right]^{1/2}.
\end{equation}
Dividing $D_z$ by $\langle \Delta t\rangle_-$ we find the group velocity below the
critical temperature.

\subsection{Numerical Results}

In Fig. 2 we plot the group velocity of a gas of sodium atoms in a box (dashed line)
and in a harmonic oscillator (dotted line) as a function of 
temperature $T$, scaled according to the critical temperature of each case. 
Density of atoms, number of atoms and trap frequencies have been taken from the data of \cite{Nature}. 
The experimental results of \cite{Nature} are seen to be broadly in agreement
with the harmonic oscillator case. The inhomogeneous spatial density of the atoms 
and its variation with temperature is the key to the understanding of the  
experimental data. The curve representing the case of free atoms shows that the 
temperature dependence entering into Eq. (\ref{susce13}) as a higher order correction has 
a negligible effect on the considered scale, and cannot be interpreted as the cause of
the behaviour observed in \cite{Nature}. \\
In Fig.3, we compare the group velocity for two 
different values of the Rabi frequency $\Omega$ coupling $|r\rangle$ to $|e\rangle$.
The behaviour for $T>T_c$ is similar to the corresponding one measured in \cite{Nature}.
The curves we obtain are however steeper, and this can be explained by considering the 
approximations made in our treatment. In our calculations we have assumed the same number
of atoms at every temperature. However, in the experiments lower temperatures are achieved 
by means of evaporative cooling. This implies that the points of the experimental curve at
higher temperatures correspond to larger numbers of atoms, and correspondingly to 
larger spatial density (for ideal gases). This leads to a smoother gradient of the group velocity 
{\it versus} the temperature than in our case. On the other hand, as the temperature 
decreases,  the effect of the interactions gets stronger causing, among other
effects, a lower density of the atoms than 
in the non-interacting case. Hence, one would expect a group velocity value
larger than the evaluated one. Albeit these considerations, 
the evaluated curve reproduces the experimental one above the critical temperature with some
agreement,
showing that the ideal gas model provides a qualitative description of the phenomenon. \\
Below the critical temperature the discrepancy between the experimental data and our
theoretical predictions is rather dramatic. This is not surprising since the 
size of the condensate is strongly affected by the effect of the interactions.
Already Ketterle and co-workers have reported that the cloud size is 
much larger in the interacting system compared to the size of the harmonic oscillator ground-state 
wave function \cite{SizeBEC}. Therefore, our evaluation can be expected to lead to smaller 
values of the group velocity than the experimental records. 
In order to illustrate this point, in the following section, 
we estimate the group velocity at $T=0$ by comparing the ideal case with the Thomas-Fermi case.\\
Finally, we discuss the measure of the group velocity in the two limiting cases 
for a section 
$S$ with radius $R\ll D_r(T)$ and with radius $R\approx D_r(T)$. This is illustrated in 
Fig. 4, 
where the same dependence of the group velocity on the temperature is evident. 
The orders of magnitude of the pairs of curves corresponding to the same set of parameters 
are comparable, showing that the behaviour observed in \cite{Nature} originates mainly
from a change in the ``average'' spatial density of the gas with temperature.

\section{Group Velocity for an Interacting Bose Gas}

In this section, we compare the group velocity value at $T=0$ in the two limits: the ideal one, 
where we consider the particles as non-interacting, and the interacting case, which we
treat in the Thomas-Fermi approximation. We estimate the group velocity using the set of parameters
of the experiment and the formula (1) of \cite{Nature}:

\begin{equation}
\label{estima}
v_g\approx\frac{\hbar c}{8\pi\omega}\frac{|\Omega|^2}{n|d_{eg}|^2}~~,
\end{equation} 
where $n$ is the density. Therefore, we need to evaluate the 
group velocity at $T=0$ by substituting into Eq. (\ref{estima}) an estimate of the 
spatial density, which we calculate here as the ratio of the total number of atoms over
the volume of the cloud. This evaluation, which corresponds to considering the density as
homogeneous, is justified on the basis of the results of Fig. 4, where it is shown that
the phenomenon observed in \cite{Nature}
is mainly dependent on the change in the density. \\
For an ideal gas in a harmonic oscillator potential at $T=0$, all the atoms are in the 
ground state, and a rough estimate of the density gives $n\approx N/(4\pi a_{0z}a_{0r}^2/3)$.
Taking $N=10^6$ Sodium atoms and 
$\nu_z=20 \times 2\pi$ Hz, $\nu_r=70 \times 2\pi$ Hz, $\Omega=0.56\gamma$, the ground state
dimensions are $a_{0z}\approx 4.7 \mu$ and $a_{0r}\approx 2.4 \mu$, and 
we obtain a density $n \approx 8\times 10^{15}$ atoms per ~cm$^3$. Thus, 
according to (\ref{estima}), 
the group velocity is  $v_g^{\text{\small{ideal}}}\approx 0.03$ m/sec. \\
For an interacting gas in the Thomas-Fermi limit,  the cloud is an ellipsoid of axes 
$2R_{TFr}$ in the radial direction and $2R_{TFz}$ in the axial direction, where $R_{TFj}$ is
the Thomas-Fermi radius:

\begin{equation}
R_{TFj}=\sqrt{\frac{2\mu}{m\nu_j^2}}\text{~~with~~}j=r,z~~,
\end{equation}
and $\mu$ is the chemical potential, defined as
\begin{equation}
\mu=\frac{\hbar\nu_{ho}}{2}\left(\frac{15 N a_S}{a_{ho}}\right)^{2/5}~~,
\end{equation}
with $a_S$ scattering length,
$\nu_{ho}=(\nu_r^2\nu_z)^{1/3}$ geometrical average of the oscillator
frequencies and $a_{ho}=\sqrt{\hbar/m\nu_{ho}}$ corresponding oscillator 
size. Taking $a_S=2.75$ nm, for the set of parameters of the experiment
the Thomas-Fermi dimensions of the cloud are $R_{TFz}\approx 47.4 \mu$ and 
$R_{TFr}\approx 13.6 \mu$~. Considering the density of atoms as homogeneous, 
we obtain $n\approx 3 \times 10^{13}$ atoms per cm$^3$.
From Eq. (\ref{estima}) we find 
for the group velocity $v_g^{\text{\small{TF}}}\approx 9$ m/sec, which is comparable with 
the value measured in \cite{Nature} for temperatures below the critical temperature.
Therefore, for $10^6$ atoms we find a difference of two orders of magnitude  
in the value of the group velocity between the ideal case and the interacting case. 
Such difference increases or decreases depending on the total number of atoms in the
trap. This estimate substantiates the inference that interactions are responsible for a lower
density, and therefore,  for a higher average group velocity of the light. 

\section{Conclusions}

We have derived an approximate analytical expression for the group velocity of a pulse
propagating through an ultracold gas which is confined in a box and by a harmonic
potential. We have shown that the results reproduce qualitatively the experimental ones
presented in \cite{Nature}. From our analysis it emerges that the
definite variation of the group velocity with the temperature of the gas is an effect
related to the variation of the spatial density of the gas. 
We see that the ideal gas model provides a qualitative description of the results for $T>T_c$. 
However, the behaviour at $T<T_c$ can be described in a satisfactory way only 
by including the interactions and the fact that the cloud is cooled by means of 
evaporative cooling. The last one has the effect of making the total 
number of atoms temperature-dependent. Such effects will be the subject of future investigations.

\section{acknowledgements}
One of us (GSA) thanks S.E. Harris, M.O. Scully and P. Meystre for
interesting discussions on the subject matter of this paper. G.M. wishes
to thank J. Schneider for many helpful comments.

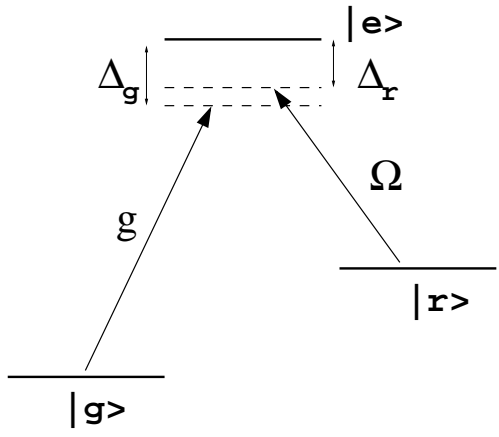
\begin{figure}
\begin{center}
\caption{Level scheme}
\end{center}
\end{figure}

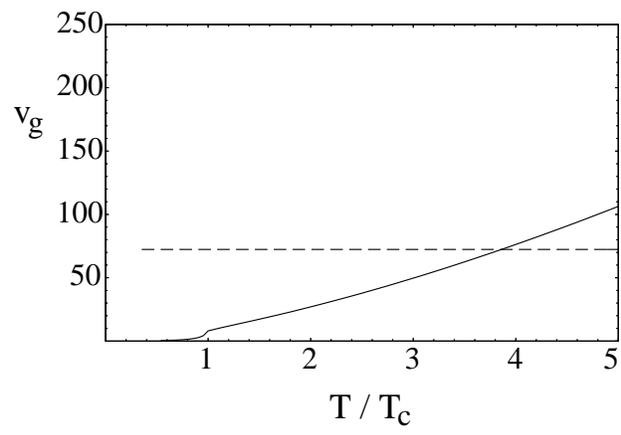
\begin{figure}
\begin{center}
\caption{Plot of the group velocity in m/sec as a function of the 
effective temperature $\theta=T/T_c$, for a gas of sodium atoms: (a)  
free bosons (dashed line) and (b) trapped bosons (dotted line). 
Here $\Omega=0.56\gamma$, $\Delta_g=\Delta_r=0$, $\Gamma_{gr}=2\pi 1000$ Hz.
For the free case: $T_c=154$ nK, $n=3.8\times 10^{12}\mbox{cm}^{-3}$; For the
harmonic potential case: $\nu_r=2\pi\times70$ Hz, $\nu_z= 2\pi\times 20$ Hz,
$T_c=432$ nK, $N=8.3\times 10^6$.}
\end{center}
\end{figure}

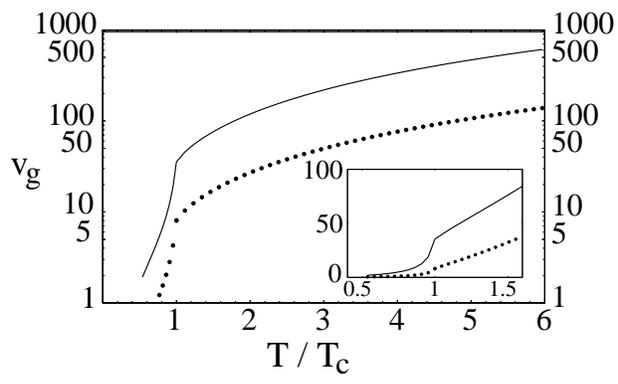
\begin{figure}
\begin{center}
\caption{Onset: Plot of the group velocity in m/sec in logarithmic scale as a function of the
effective temperature $\theta=T/T_c$ for a gas of sodium atoms as in [7]. 
The upper curve corresponds to $\Omega=1.2\gamma$, whereas the lower curve 
corresponds to $\Omega=0.56\gamma$. Here, $T_c=432$ nK, $N=8.3\times 10^6$, 
$\Gamma_{gr}=2\pi\times 1000$ Hz,
$\nu_r=2\pi\times70$ Hz, $\nu_z=2\pi\times 20$ Hz and $\Delta_g=\Delta_r=0.$ Inset: Plot of the low 
temperature behaviour of the corresponding curves in linear scale. The radius of the section $S$ is
R=15 $\mu$.}
\end{center}
\end{figure}

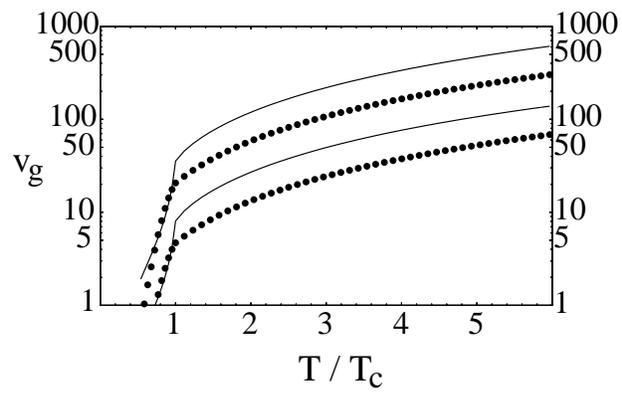
\begin{figure}
\begin{center}
\caption{Calculations with two different radii for the section $S$: $R=15\mu$
(solid lines) and $R=\sqrt{K_BT/m\nu_r^2}$ (dotted lines). The two bottom (top) curves
correspond to $\Omega =0.56\gamma~ (\Omega=1.2\gamma)$. All the other parameters
are reported in the caption of Fig.3.}
\end{center}
\end{figure}
\end{document}